\documentclass[11pt]{article}  
 \pdfoutput=1  
\usepackage{jheppub}
\usepackage{titlesec}
\usepackage{tikz, amsmath} 
\usetikzlibrary{calc} 
\usepackage{bbm}
\usepackage{xcolor} 
\usepackage{dsfont}
\usepackage{subfigure} 
\usepackage{hyperref} 
\usepackage{microtype}
\usepackage{caption} 
\usepackage{amsmath}
\usepackage{float} 
\usepackage{graphics}
\usepackage{psfrag}
\usepackage{color}
\usepackage{slashed} 
\usepackage{subfigure}
\usepackage{graphicx}
\usepackage{array,multirow}
\usepackage{amsfonts}
\usepackage{amssymb}
\usepackage{mathtools}
\usepackage{amsmath}
\usepackage{amsfonts}
\usepackage{mathrsfs}
\usepackage{multirow}
\usepackage{array}
\usepackage{textcomp} 
\usepackage{phaistos} 
\usepackage[utf8]{inputenc}
\usepackage{pifont} 
\usepackage{pgf}

\usepackage{subfigure}
\usepackage{multirow}

\newcommand{\beq}{\begin{eqnarray}}
\newcommand{\eeq}{\end{eqnarray}}

\newcommand{\bmp}{\noindent\begin{minipage}{16cm}}
\newcommand{\emp}{\end{minipage}\vskip 7mm} 

\usepackage{dcolumn}
\usepackage{bm}
\usepackage{slashed} 

\usepackage{epsfig}

\newcommand{\bea}{\begin{eqnarray}}
\newcommand{\eea}{\end{eqnarray}}

\newcommand{\ba}{\begin{eqnarray}}
\newcommand{\ea}{\end{eqnarray}}

\newcommand{\de}{\partial} 

\newcommand{\drawsquare}[2]{\hbox{%
\rule{#2pt}{#1pt}\hskip-#2pt
\rule{#1pt}{#2pt}\hskip-#1pt
\rule[#1pt]{#1pt}{#2pt}}\rule[#1pt]{#2pt}{#2pt}\hskip-#2pt
\rule{#2pt}{#1pt}}

\newcommand{\Yfund}{\raisebox{-.5pt}{\drawsquare{6.5}{0.4}}}

\title{Effective Lagrangian and Stability Analysis in Warped Space}  

\author{Haiying Cai}
\emailAdd{hcai@korea.ac.kr}
\affiliation{ Department of Physics, Korea University, Seoul 136-713, Korea}  

\abstract {In the warped space model,  the inter-brane distance can be stabilized by the Goldberger-Wise mechanism.  Of  particular importance, the stabilization potential  calls for a proper identification of the dynamical degree of freedom. In this paper, we provided a complete calculation of the effective Lagrangian till the quadratic order that is generic for the Randall-Sundrum model and its $N$-brane $(N \geq 4)$ extensions.  By  applying the variation principle to a specific perturbation field,  we  derived the  equations of motion and  orthogonal conditions for decoupling the graviton. This approach is demonstrated to be equivalent to the analysis using  the linearized Einstein equation. Our derivation clarifies that in the $N$-brane set up,  just one degree of freedom for the radion field is dynamical, with the other   modes  eliminated by the gauge fixings. Thus we can directly generalize the GW stabilization to the $N$-brane model in a way  similar to the RS1 scenario. }

\begin{document}
\maketitle

\section{Introduction} 
The original Randall-Sundrum (RS) model with 2 branes at the orbifold fixed points~\cite{Randall:1999vf, Randall:1999ee} was proposed to address the hierarchy problem using a warped factor. Also the localization of gravity near the UV brane naturally explains the weakness of  coupling at the large distance.  An interesting generalization of one slice of Anti-de-sitter (AdS) space is to build RS-like models with extra branes~\cite{Kogan:1999wc, Kogan:2000xc, Kogan:2001wp}. Such construction is attractive for the existence of  ultra-light massive gravitons and potential new phenomenology. A general warped multibrane  model was described in~\cite{Kogan:2001qx}, where two non-fixed point branes were added given that the bulk cosmology constants are different in two spatial regions. For the RS1 model, the inter-brane distance has to be stabilized  by  the Goldberger-Wise (GW) mechanism~\cite{Goldberger:1999uk,  Goldberger:1999un}, that requires at least  a single bulk scalar  coupling to the gravitons.  As a result, the fluctuation of the bulk scalar becomes entangled with the metric modulus field.  Meanwhile the  effective potential of the scalar develops a minimum after  an  integration over the fifth  dimension so that the radion obtains a mass.   

Recently we have proved that this stabilization mechanism can be  generalized  to a multibrane set-up in a  straightforward manner~\cite{Cai:2021mrw}. In that paper, with the addition of a new perturbation $\epsilon$ in the metric, we derived  the linearized Einstein equation in a multibrane RS model with the junction conditions matched at all the branes. However the preliminary analysis  shows that the perturbation $\epsilon$  simply plays the role of gauge fixing. Following the strategy of \cite{Csaki:2000zn}, we can solve  a single equation of motion (EOM)   as an eigenvalue problem in the limit of stiff brane potentials and find that with  a small back-reaction the  mass of radion  is significantly suppressed compared to its KK excitations \cite{Cai:2021mrw}. In this paper, in order to strengthen the argument of dynamical degree of freedom, we  expanded the 5d action into the quadratic order of perturbations. Despite of the complexity,  by applying the variation principle to the effective Lagrangian, we derived  the same EOMs and orthogonal conditions as from the  Einstein equation. Note that  the impact of  perturbation $\epsilon$ on the scalar EOM can only be explored in the framework of effective Lagrangian.  Of particular interest is that  the dependence of $\epsilon$ in the effective Lagrangian  can be  eliminated after imposing the orthogonal conditions (gauge fixings). This constitutes a stronger demonstration  that  a unique  radion field with its profile $F \propto e^{2A} $ at the zeroth order  is  the legitimate solution to the Einstein equation in a stabilized $N$-brane model. 

As a consistency check of stability, we further examined the tadpole behavior of  the lowest mode of radion-scalar system after the GW stabilization. The result shows that the linear terms of  $\epsilon$ and bulk scalar fluctuation  are removed by the radion EOMs and background equations, while the remaining  term related to the 5d profile is  automatically zero at the leading order.

\section{5d model and diffeomorphism} \label{review}
\begin{figure}
	\centering 	
	{\includegraphics[height=4.0 cm,  
		width=7.0 cm]{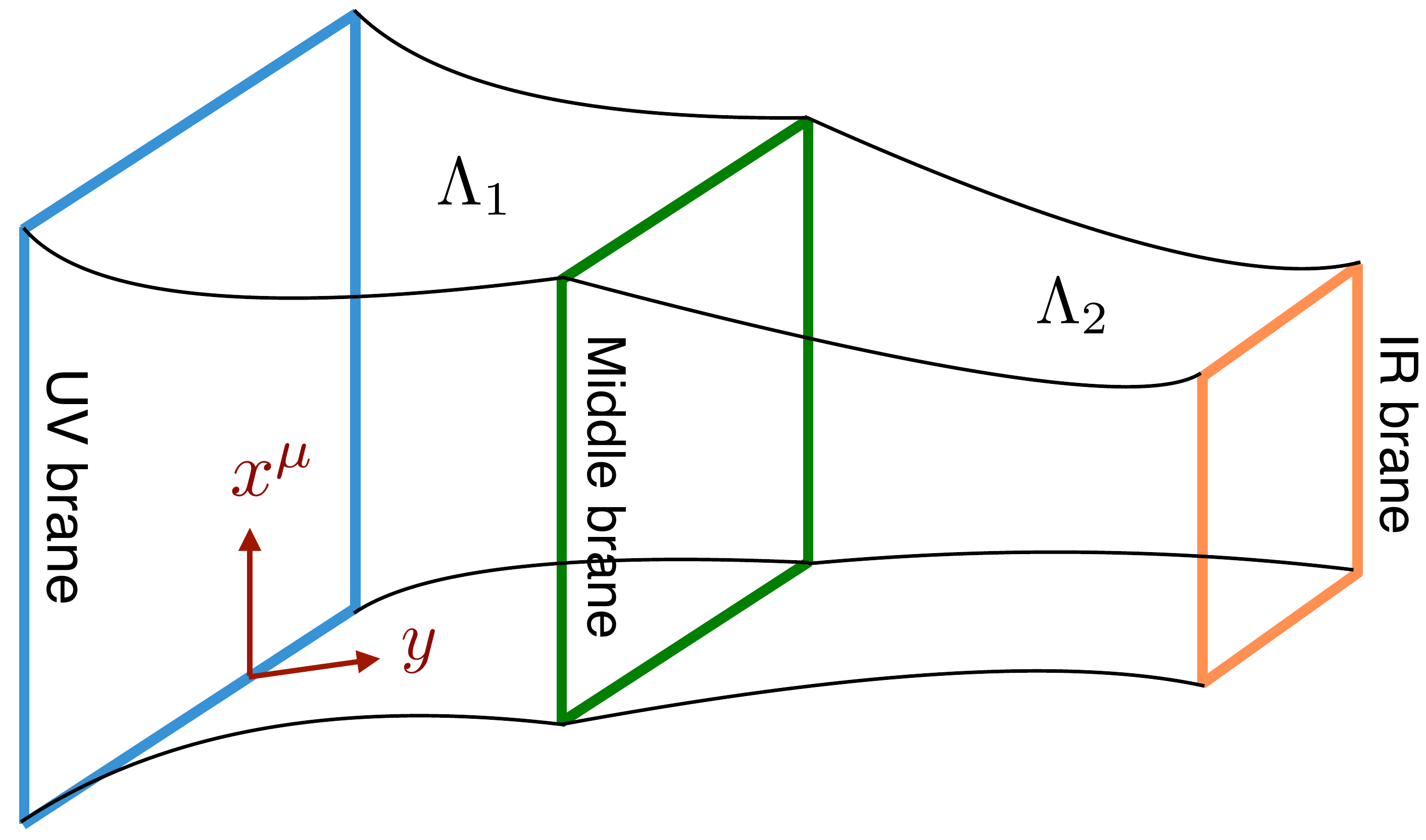}} 
	\caption{The $4$-brane  RS model  visualized in the range of $0 < |y| < L$. Due to the orbifold symmetry,  there are two intermediate branes located at $y=\pm r$.} \label{fig: multibrane}
\end{figure}

We start with a brief review of  $N$-brane ($N \in$ even integer) model, considering the 5d action with the graviton minimally coupling to a single bulk  scalar field:
\beq
& - & \frac{1}{2 \kappa^2} \int d^5 x \sqrt{g} \, {\cal R} + 
\int d^5 x \sqrt{g}  \Big( \frac{1}{2} g^{I J}  \partial_I \phi \partial_J \phi
-  V(\phi) \Big)  \nonumber \\ &-& \int d^5x\frac{\sqrt{g}}{\sqrt{-g_{55}}}\sum_{i} \lambda_{i}(\phi) \delta(y - y_i) \,, \label{Act} 
\eeq
where $\mathcal{R}$ is the Ricci scalar and the Latin indices $I,J$ run over $(\mu, 5)$,  with the Greek one designated for 4d Minkowski space  $\mu =0,\cdots 3$ and $y$ being the coordinator of  extra dimension. The  $\kappa^2 \equiv 1/ (2 M_5^3) $ is related to the 5d Planck mass.  Note that the orbifold symmetry is imposed for the Lagrangian where all the fields satisfy $Q(y) = Q(-y)$, thus the integration of $y$ is  conducted in the region of $y \in [-L, L]$ that is equivalent to a  circle $S^1$ under the diffeomorphism. In Eq.(\ref{Act}),   the first line contains the  Einstein-Hilbert action  and the bulk scalar action that is responsible for the GW mechanism. While the second line is composed of the brane terms determined by the  {\it{jump}}~\footnote{The {\it jump} of a given quantity $Q$ cross the brane located at $y= y_i$ is defined as  $[Q]|_{y= y_i} \equiv  Q(y_i+\varepsilon) - Q(y_i-\varepsilon)|_{\varepsilon \to 0}$ }  of the derivative fields as well as the brane potentials of $\phi$. With the appropriate  potentials $V(\phi)$ and $\lambda_i(\phi)$, $i = 1, \cdots N$, the bulk scalar will develop a  VEV i.e. $\phi(x,y) = \phi_0(y) + \varphi(x,y) $, so that the radion field is stabilized. In an $N$-brane set up,  the fifth dimension can be divided into $N/2$  subregions with different curvatures $k_\alpha^2  = -  \frac{\kappa^2}{6} \Lambda_\alpha $ , $\alpha = 1, \cdots \frac{N}{2}$ , where  $\Lambda_\alpha$ is the cosmology constant in each subregion.  Henceforth besides the UV and IR branes at the fixed points of $y= \{ 0, L\}$ like the  RS1 model, $N-2$ copies of intermediate branes (not dynamical as later proven) arise  at $y= \pm r_a $ ($0  < r_{a} < L$) with energy densities to match the junction conditions of the metric.  As a concrete example, the $4$-brane RS model is displayed in  Figure \ref{fig: multibrane}.

The general metric ansatz  on an $S^1/Z_2$ orbifold  that can decouple  the transverse graviton  from the physical radion field is~\cite{Charmousis:1999rg, Csaki:2000zn}:
\beq
d s^2  &=&   e^{-2 A (y) - 2 F(x, y) }\left[ \eta_{\mu \nu} + 2 \epsilon(y) \partial_{\mu} \partial_{\nu} f (x) \right. \nonumber \\ 
&+& \left. h_{\mu \nu}(x,y)\right] d x^\mu d x^\nu - \left[1+ G(x, y) \right]^2 dy^2 \,, \label{metric} 
\eeq
with $A(y)$ being  the background metric. Among the perturbations,  $h_{\mu \nu}$ is a symmetric  tensor standing for the graviton.  While $F, G, \epsilon f(x)$ plus the scalar perturbation $\varphi$ are not independent, they  all belong to  the radion excitation. Compared with the previous paper~\cite{Cai:2021mrw},  the definitions of $F, G $ absorb the $f(x)$ that is factorized outside for $\epsilon(y)$. We would like to mention that Eq.(\ref{metric}) is enforced an  implicit constraint $g_{\mu 5}=0$  otherwise there is one more scalar mode that could be removed by the gauge fixing~\cite{Gherghetta:2011rr}. 

Before the radius stabilization, it is inspiring to investigate the transformation property of these metric perturbations under a class of infinitesimal coordinate shift
$X^I \to X^I + \xi^I(X)$. As a result, the metric  transforms accordingly:
\beq 
\delta g_{IJ} = - \xi^K \, \partial_K  g_{IJ}^{(0)} - \partial_I \xi^K \,  g_{K J}^{(0)} - \partial_J \xi^K \, g_{IK}^{(0)} \,. \label{dG}
\eeq
Note that the diffeomorphism  symmetry  keeps  the Einstein-Hilbert action to be  invariant.  Since the Ricci scalar is purely constructed by the metric,  the diffeomorphism  will  retain  the metric in its original structure  after the appropriate field redefinition. This requires the transformation to be of the specific form~\cite{Kogan:2001qx,Gherghetta:2011rr}:
\beq
\xi^\mu(x,y) &=& \hat{\xi}^\mu(x)  + \, \eta^{\mu \nu}  \partial_\nu   \zeta(x, y) \,,  \quad    \xi^5(x,y) = e^{-2A} \, \zeta^\prime(x, y) \,. \label{xi}
\eeq
where the prime denotes the derivative $\partial_5 \equiv \de / \de y$. Due to the presence of the brane terms, the fifth coordinate shift is  subject to a constraint  $\zeta^\prime(x, y)|_{y = \{0, \pm r_a, L\}} =0$, $a = 1, \cdots \frac{N}{2}-1$.  Substituting  the $\xi^K$ of Eq.(\ref{dG}) in terms of Eq.(\ref{xi}),  one can extract the transformation rules for all  component fields in the metric expansion: 
\beq
&& \delta h_{\mu \nu} = - \partial_{\mu} \hat{\xi}_{\nu} - \partial_\nu \hat{\xi}_\mu \nonumber\\
&&\delta F = - A'  \zeta'  e^{-2A} \nonumber \\
&&\delta G =  - \left( \zeta'' - 2 A' \zeta' \right)  e^{-2A} \nonumber\\
&& \delta \epsilon f(x) =- \zeta  \,. \label{transform}
\eeq
As we can see, $\hat{\xi}^{\mu}(x) $ represents the usual 4d diffeomorphism under which $h_{\mu \nu}$ transforms  as a  spin-2 tensor. For the spin-0 modes $G$, $F$ and $\epsilon f(x)$, their transformations are  fixed by  the metric $A$ and the  parameter $\zeta$.  In particular,  the last ansatz in Eq.(\ref{transform}) indicates that without stabilization the $\epsilon f(x) $ can  be eliminated  as  a gauge fixing by choosing $\zeta = \epsilon f(x)$.

In general, the  system of metric and scalar perturbations can be studied using the variation principle since the physical path  evolves along the one minimizing the action. Varying of the action (\ref{Act}) with respect to the 5d metric $g_{IJ}$ will give the  Einstein Equation:
 \beq
 R_{IJ}  = \kappa^2 \left(T_{IJ} -\frac{1}{3} g_{IJ} T^m_m\right) = \kappa^2 \tilde{T}_{IJ}\,, \label{EE}
\eeq
 with $T_{IJ} = 2 \delta \left(  \sqrt{g}\, \mathcal{L}_{\phi}\right) / \left( \sqrt{g} \, \delta g^{IJ} \right) $ being the energy-momentum tensor.  Similarly by minimizing the scalar action with  respect to $\phi$, one can derive the scalar EOM that is  not included in Eq.(\ref{EE}).
Grouping  the  zeroth order of Eq.(\ref{EE}) with the equation of  $\phi_0$ together,  the background  (BG) equations  in an $N$-brane RS model are:
\beq
&&\phi_0 '' = 4 A' \phi_0 ' + \frac{\partial V(\phi_0)}{\partial \phi}+ 
\sum_{i} \frac{\partial \lambda_i(\phi_0)}{\partial \phi} \delta(y-y_i)\,,  \label{BG0} \\
&& A'' = \frac{ \kappa^2 }{3} \phi'^2_0 + \frac{\kappa^2 }{3}
\sum_i \lambda_i(\phi_0) \delta(y-y_i) \,, \label{BG1}
\\
&& A'^2 = \frac{\kappa^2 {\phi_0}'^2}{12} 
- \frac{\kappa^2}{6} V(\phi_0) \,.
\label{BG2}
\eeq 
The delta functions  in. Eq.(\ref{BG0}) and Eq.(\ref{BG1}) signal the discontinuity of $\phi'_0$ and $A'$ at the boundaries. Note that the three BG equations are not independent due to the Bianchi identity.  First let us take a $\partial_5$ operation on Eq.(\ref{BG2}). Then by  inserting Eq.(\ref{BG1}) to the differentiated ansatz, we will arrive the scalar  BG equation Eq.(\ref{BG0}). The coupled BG equations can be analytically solved in terms of a single super-potential function $W(\phi)$ \cite{DeWolfe:1999cp, Behrndt:1999kz}, with the solutions  written as:
\beq
&&\phi'_0 = \frac{1}{2}\frac{ \partial W(\phi_0)}{\partial \phi_0} \,, \quad \quad \quad  A' = \frac{\kappa^2}{6} W(\phi_0) \,, \nonumber \\ 
&&V(\phi) =  \frac{1}{8} \left(\frac{ \partial W(\phi)}{\partial \phi}\right)^2 -\frac{\kappa^2}{6} W(\phi)^2 \,. \label{sup}
\eeq  
In this approach the back-reaction of the bulk scalar on the metric is automatically included. 
It is well known that  the linear  expansion of Einstein equation (\ref{EE})  gives the EOMs of graviton ($h_{\mu \nu}$) and  radion fields ($F$ and $G$)~\cite{Cai:2021mrw}. However, one must  expand the 5d action till the quadratic order,  so that the scalar EOM (modified by a shift of $ \phi'_0 \epsilon' \Yfund f (x)$ in the metric of Eq.({\ref{metric}}))  can be obtained by the variation principle. Furthermore,  an effective Lagrangian with explicit  kinetic terms  is indispensable for  phenomenology study, thus necessary for working out. In the next section, we will demonstrate  that all the  EOMs  and  orthogonal conditions can be derived with more clarity  in  the formalism of  effective Lagrangian. 

\section{The Effective Lagrangian} \label{sec:EFT}
The 5d action Eq.(\ref{Act}) can be expanded in terms of  the metric and scalar perturbations.  We will postpone the discussion of tadpole term in next section.  The effective Lagrangian  at the quadratic order is:
\beq     
\mathcal{L}_{eff} & = & \int dy  \frac{e^{-2A}}{2\kappa^2} \Big\{  \frac{e^{-2A}}{4} \left[(\partial_5 h)^2 - \partial_5 h_{\mu \nu} \, \partial_5 h^{\mu \nu} \right] - {\mathcal L}_{FP}  \nonumber \\   
&-&  \left[ G- 2 F - e^{2A}  \partial_5\left( \epsilon'  f(x) e^{-4A}\right) \right] \left( \partial_{\mu} \partial_\nu h^{\mu \nu} - \Yfund h\right) 
\nonumber \\ 
&-& 3 e^{- 2A} \left[ F'  -A' G  - \frac{\kappa^2}{3} \phi'_0 \varphi \right] \partial_5 h  -    \kappa^2 e^{-2A}  \mathcal{L}_{5m} \nonumber \\
& + &  \kappa^2  \partial_\mu \varphi \partial^\mu \varphi  -  6 \, \Big[ \partial_\mu F \partial^{\mu} (F-G) \nonumber \\ &-&  e^{-2 A} \epsilon'  \partial_\mu \left[F'  - A' G  - \frac{\kappa^2}{3} \phi'_0 \varphi \right] \partial^\mu f(x) \Big]  \Big \} 
\label{eff}   
\eeq
where  $\mathcal{L}_{FP}$ is the Fierz-Pauli Lagrangian in Eq.(\ref{FP}). Note that  all the terms with  graviton and the kinetic term of  radion are derived in Eqs.(\ref{grav},\ref{mix},\ref{rad})  in  Appendix~\ref{Appendix1}. Specifically  $\mathcal{L}_{5m}$ contains the quadratic terms of radion without the $\partial_\mu$ operator. We define that:
\beq
&\mathcal{L}_{5m}&  = \frac{4}{\kappa^2}  \Big[  2( G + 4F) F'' -  10 (G + 4F) A' F' \nonumber \\  &+& 5 F'^2 -4 (2 F + G) A' G'  + 2F' G'  \nonumber \\ &+& G^2 (5 A'^2 - 2 A'') + 16 F^2 (A'^2 -A'')  \nonumber \\ &+& 8 F G (4 A'^2 - A'') \Big]   +  \varphi'^2+   \left[ G^2 + 16 F^2\right]  \phi'^2_0  \nonumber  \\ &-&  2 (G+ 4F) \phi'_0 \varphi'  +
 \left[ 2 (G- 4F)  \frac{\partial V}{\partial \phi_0} \varphi  +   \frac{\partial^2 V}{\partial \phi_0^2} \varphi^2 \right]\nonumber \\
&+&  \sum_i \Big[ 8 F \left(2 F  \lambda_i -  \frac{\partial \lambda_i}{ \partial \phi_0}\varphi  \right)  +  \frac{\partial^2 \lambda_i}{\partial \phi_0^2}\varphi^2  \Big] \delta(y-y_i) \, \label{mass1}
\eeq
For convenience,  the terms  in $\mathcal{L}_{5m}$ will be classified. In fact the expansions of Einstein-Hilbert action $S_{EH} \sim \int d^5 x \sqrt{g} \,\mathcal{R}$ are  put in the first square parenthesis. While the remaining items are  from other origins that are not absorbable to the kinetic term.  As we emphasized earlier that  the $y$-integration path is along a circle $S^1$, hence any total  differentiation term can be set to be zero.  Using the tricks of partial integration,   the terms  in $\mathcal{L}_{5m}$  from the Ricci scalar can be  recasted into a concise expression:
\beq
- \frac{12}{\kappa^2}  \Big[ F'^2 +  G^2 A'^2 - 2 F' G A'   + 4 F^2  A'' \Big]  
\eeq
where the  $A''$ will be substituted  by  the expression in Eq.(\ref{BG1}). Then in Eq.(\ref{mass1})  the two terms in the forms of  $F^2 \phi'^2_0$ and $ \lambda_i F^2 \delta(y-y_i)$ are precisely  eliminated. Hence  $\mathcal{L}_{5m}$ is further simplified to be:
\beq
\mathcal{L}_{5m} & = &- \frac{12}{\kappa^2}  \Big[ F'^2 +  G^2 A'^2 - 2 F' G A'   \Big]  +     \varphi'^2 + G^2  \phi'^2_0 -  2 (G+ 4F) \phi'_0 \varphi' \nonumber \\ &+&   \left[ 2 (G- 4F)  \frac{\partial V}{\partial \phi_0} \varphi  +   \frac{\partial^2 V}{\partial \phi_0^2} \varphi^2 \right] -  \sum_i \Big[  8  \frac{\partial \lambda_i}{ \partial \phi_0} F \varphi  -  \frac{\partial^2 \lambda_i}{\partial \phi_0^2}\varphi^2  \Big] \delta(y-y_i) \, \label{mass2}
\eeq
Eq.(\ref{mass2}) shows that only  $G^2 \phi'^2_0$ survives  after the stabilization while  other metric expansions  proportional to $\phi'^2_0$ are all cancelled.
  
Now we are ready to practice  the variation principle without  imposing any gauge fixing in advance. One can vary the effective Lagrangian Eq.(\ref{eff}) with respect to $ G, F,  \varphi $. This will give 3 equations of motion: 
\beq
&& ~~\phi'_0 \varphi'   - G\phi'^2_0 - \frac{\partial V}{\partial \phi_0}\varphi  \nonumber \\ 
&& =  \frac{3}{\kappa^2} \left[4 A' ( F' - A' G)  +   \Yfund \left(F e^{2A}  -A' \epsilon'  f(x) \right) \right] \,, \label{eom1}\\ 
&& ~~ \left( \phi'_0 \varphi' +  \frac{\partial V}{\partial \phi_0}\varphi  \right) 
+  \sum_i \left( \lambda_i G + \frac{\partial \lambda_i}{ \partial \phi_0}  \varphi \right) \delta(y-y_i)  \nonumber\\
&& =  \frac{3}{\kappa^2} \left[  F'' - G' A' - 4 A' F'   \right] - 2 GV  \nonumber   \\
&& +  \frac{3}{4 \kappa^2} e^{2A} \Yfund  \Big( G - 2 F  - e^{-2A} \left[ \epsilon'' - 4 A'\epsilon'\right] f(x) \Big)\,, \label{eom2} \\
&& (G' + 4F' ) \phi'_0 +   4 A' \varphi'   + \sum_i \left (  \frac{\partial \lambda_i}{ \partial \phi_0} G + \frac{\partial^2 \lambda_i}{ \partial \phi_0^2} \varphi  \right) \delta(y-y_i) \nonumber \\
&&= \varphi''  - \left( 2   \frac{\partial V}{\partial \phi_0} G  + \frac{\partial^2 V}{\partial \phi_0^2} \varphi \right) -  \Yfund  \Big( \varphi  e^{2A} -\phi'_0 \epsilon' f(x) \Big) \,.  \label{eom3}
\eeq
In addition by requiring  no  mixing between the graviton and  radion,  the 3rd and 4th terms in Eq.(\ref{eff}) lead to two orthogonal conditions:
\beq
&& F'  - A' G  - \frac{\kappa^2}{3} \phi'_0 \varphi  = 0  \,,  \label{orth1}  \\
&& G - 2 F  - e^{-2A} \left[ \epsilon'' - 4 A'\epsilon'\right] f(x) =0  \,.  \label{orth2}  
\eeq  
Eqs.(\ref{orth1}-\ref{orth2}) are precisely the transverse and traceless gauge fixings derived in \cite{Cai:2021mrw} that can decouple the graviton. Note that the appearance of $\epsilon'$ in EOMs (\ref{eom1}-\ref{eom3}) comes purely from the  variation of a term $- \frac{3}{\kappa^2} \int dy e^{-4 A} \epsilon'  \partial_\mu \left[F'  - A' G  - \frac{\kappa^2}{3} \phi'_0 \varphi \right] \partial^\mu f(x)$, i.e. $\epsilon'$ times the first orthogonal condition\footnote{$\epsilon' f(x)$ behaves like a Lagrange multiplier. Varying Eq.(\ref{eff}) with respect to $\epsilon' f(x)$ gives back Eq.(\ref{orth1}).}.

In the following, we will  prove that the formalism of effective Lagrangian  is equivalent to the  linearized Einstein equation, by providing the same set of correlated EOMs.  Firstly we can  identify that Eq.(\ref{eom3}) is just the  EOM  of bulk scalar. Compared with the case without $\epsilon$ perturbation,  the scalar EOM  is modified with a shift of $\phi'_0 \epsilon' \Yfund f(x) $.  Next using Eqs.(\ref{BG0}-\ref{BG1}) and Eq.(\ref{orth1}), the first EOM (\ref{eom1}) can be transformed to be:
\beq
&&\Yfund \left[F e^{2A}  -A' \epsilon'  f(x) \right] +  A'' G  - \frac{\kappa^2}{3} \left( \phi'_0 \varphi'  - \phi''_0 \varphi \right) \nonumber \\
&=& \frac{\kappa^2}{3} \sum_i \left( \lambda_i(\phi_0) G + \frac{\partial \lambda_i(\phi_0) }{\partial \phi}  \varphi \right) \delta(y-y_i) \, \label{eom1a}
\eeq
Then after taking the differentiation of Eq.(\ref{orth1}), i.e.
\beq
\frac{\kappa^2}{3} \phi''_0 \varphi = \left( F'' - A' G' -A'' G\right) -\frac{\kappa^2}{3} \phi'_0 \varphi'  \, \label{dorth}
\eeq
we'll  insert Eq.(\ref{dorth}) into Eq.(\ref{eom1a}) and obtain: 
\beq
&&\Yfund \left[F e^{2A}  -A' \epsilon'  f(x) \right]  +\left( F'' - A'G' \right) - \frac{2\kappa^2}{3}  \phi'_0 \varphi'  \nonumber \\
&=& \frac{\kappa^2}{3} \sum_i \left( \lambda_i(\phi_0) G + \frac{\partial \lambda_i(\phi_0) }{\partial \phi} \varphi  \right) \delta(y-y_i) \, \label{eom}
\eeq
that  is exactly the EOM of radion derived in \cite{Cai:2021mrw}.  We would like to remark that an exact correspondence can be established between the first two EOMs and the Einstein equation Eq.(\ref{EE}). In fact one can find that Eq.(\ref{eom1}) is  from the  assembling of $ \frac{1}{\kappa^2}\left[e^{2 A} R_{\mu \nu} /\eta_{\mu \nu} + R_{55} \right] $, while  Eq.(\ref{eom2})  corresponds to the combination  of $ -\frac{1}{2 \kappa^2}\left[ 2 e^{2 A} R_{\mu \nu} /\eta_{\mu \nu} - R_{55} \right]$.  The correlation between the $(\mu \nu)$ and $(55)$ components of Eq.(\ref{EE}) is demonstrated in Appendix~\ref{Appendix3}. Hence one can obtain Eq.(\ref{eom2})  directly from Eq.(\ref{eom1}) after some lengthy algebra,  with the necessity to remove the last term in  Eq.(\ref{eom2}) by enforcing the second orthogonal condition.

Finally  the scalar EOM (\ref{eom3}) is not independent to the Einstein equation. Dropping out the brane terms,  one can first construct  an ansatz of $ \partial_5 \left[e^{-2A}  \mbox{Eq.(\ref{eom})}\right]$, then subtract it with $e^{-2A} A' \left[R_{55} - \kappa^2 \tilde{T}_{55}\right] =0 $. The resulting equation containing a term of $\Yfund (F' -A' G)$ is  actually the scalar EOM (\ref{eom3}) times $\kappa^2 \phi'_0 e^{-2A} /3 $  (see Appendix~\ref{Appendix3}).  

Therefore  although we start with 4 radion related scalars,  only  a single  EOM Eq.(\ref{eom}) plus the two gauge fixings Eqs.(\ref{orth1}-\ref{orth2}) are independent ones, indicating that one perturbation  is not dynamical.  Recalling the diffeomorphism in Eq.(\ref{transform}), before stabilization one can  always set  $\zeta =  \epsilon f(x)$  to eliminate the arbitrariness  in the bulk. Simultaneously  $\epsilon'$ will inherit  the zeros of $\zeta'$  at all the branes. Is this  gauge fixing  still operative in the presence of  stabilization?  With the constraint $\epsilon'|_{y_i}  =  \epsilon''|_{y_i}  = 0$, $y_i = \{0, \pm r_a, L\}$, $a = 1, \cdots \frac{N}{2}-1$,  it is  viable to conduct the field redefinition according to a spurious symmetry,      
\beq
\tilde{F} &=& F - A' \epsilon' f(x) e^{-2A} \nonumber \\
\tilde{G} &=& G - \left( \epsilon'' - 2 A' \epsilon' \right) f(x) e^{-2A} \nonumber \\
\tilde{\varphi} &=& \varphi - \phi'_0 \epsilon' f(x) e^{-2A} \, \label{zeta}    
\eeq
so that the $\epsilon'$ can be fully removed from the EOMs (\ref{eom1}-\ref{eom3}) and Eqs.(\ref{orth1}-\ref{orth2}) become $\tilde{F}'  - A' \tilde{G}  = \frac{\kappa^2}{3} \phi'_0 \tilde{\varphi} $ and  $\tilde{G} = 2 \tilde{F}$. But  the justification of this transformation should be investigated. Inspecting the 5d action first,  the kinetic term  is shifted by:
\beq
\frac{3}{\kappa^2} \int dx^5   \Big(e^{-4A} \partial_\mu \tilde{F} \left[ \epsilon'' - 2 A' \epsilon' \right]   + \frac{A'}{2}  \frac{d}{dy} \left[ \epsilon'^2 e^{-6A}\right]  \partial_\mu f(x)\Big)  \partial^\mu f(x)  + \Box (\delta \varphi) ~\mbox{terms} \, \label{deltaS} 
\eeq 
which depends on the bulk value of $\epsilon'(y)$ for $\phi'_0 \neq 0$. This implies that the $\zeta$-symmetry (relic of 5d diffeomorphism) in Eq.(\ref{transform})  is spontaneously broken if the GW scalar develops a $y$-dependent VEV.  Actually  as  $\varphi$ must be invariant, $\delta \varphi = 0$ in Eq.(\ref{zeta}) will force $\phi'_0 \, \epsilon' = 0$ to preserve the 4d  Poincar\'e invariance,  otherwise ambiguity will enter in the radion kinetic term. Thus two cases are  permitted according to the symmetry principle: 
\begin{itemize}
\item[(a)] For $\phi'_0(y) \neq 0$, one will get $\epsilon'(y) =0$, i.e. $\epsilon=$ {\it constant} is a pure  gauge in Eq.(\ref{metric}) without  impact on the dynamics. 
This option can provide a static radion solution, while signals  the  breaking of $\zeta$-symmetry. 

\item[(b)] For $\phi'_0(y) =0 $,  $\epsilon'(y) \neq 0$ is allowed, that preserves the 5d diffeomorphism if $\epsilon'|_{y_i} =0$.  With $\phi'_0=0$,  one can also relax the BC to be $\epsilon' (r_a) \neq 0$ and Eq.(\ref{deltaS}) is  a nonzero surface term because of $\tilde{F} \sim e^{2A}$ and $A' \sim$ {\it constant}.  However the second option offers no radion stabilization. 

\end{itemize}
Consequently  case (a) is the correct option for  a stabilized radion.   In fact,  simply imposing the two orthogonal conditions (\ref{orth1}-\ref{orth2}) on the effective Lagrangian Eq.(\ref{eff}), we find that  no any $\epsilon' f(x)$ could remain.  This leaves the  kinetic terms of the graviton and radion to be:
\beq     
\mathcal{L}_{kin} & = & -  \frac{1}{2\kappa^2} \int dy  \, e^{-2A} \Big\{   {\mathcal L}_{FP}   -  \kappa^2  \partial_\mu \varphi \partial^\mu \varphi  
 + 6 \, \Big[ \partial_\mu F \partial^{\mu} (F-G)\Big]   \Big \} 
\label{kinetic}   
\eeq 
Therefore the solvable radion EOM (respecting 4d diffeomorphism) should be Eq.(\ref{eom}) gauged with $\epsilon' =0$ and $G = 2 F$.  Notice that this property deduced from the symmetry principle is valid for any $N$-brane ($N \geq 2$)  RS model.

\section{The tadpole term}
Now we will discuss the tadpole term that is  the linear order expansion of Eq.(\ref{Act}), since a sizable tadpole might disturb the radion EOM.  In the literature the radion tadpole is linked to one of the sum rules in the brane worlds~\cite{Gibbons:2000tf},  derived from the background Einstein equation in a spatially periodic extra dimension.  Assuming  the internal curvature of Minkowski space is zero and  $F \sim e^{2A}$,  the paper~\cite{Papazoglou:2001ed} claimed that the  coefficient of the linear term $F$  is proportional to
\beq
 \oint dy e^{-2A} \left[  \phi'^2_0 + 2V(\phi_0) + 2 \sum_i \lambda_i \delta(y- y_i) \right] =  0 \, \label{sumrule}
\eeq
that is zero and can be immediately verified  using Eq.(\ref{sup}) (see Appendix~\ref{Appendix2}).   However the weakness in that argument is the  scalar perturbation $\varphi$  is fully ignored. Indeed  for $\phi_0 =0$ and $V(\phi_0) = -\frac{6}{\kappa^2} A'^2$ ,  the tadpole of radion field is bound to vanish due to Eq.(\ref{sumrule})~\cite{Cai:2021mrw}. In this section we will  provide a rigorous calculation for the radion tadpole, that does not  align to  Eq.(\ref{sumrule})  after  the GW  stabilization. 

The derivation is proceeded by employing  the Einstein equation Eq.(\ref{EE})  to  transform the  Ricci scalar $\mathcal{R}$  into the forms of  $V(\phi_0)$ or $\lambda_i (\phi_0)$ and their first derivatives with respect to $\phi_0$. From the compact expression of the modified energy momentum tensor $\tilde{T}_{IJ} = T_{IJ} -\frac{1}{3} g_{IJ} T^m_m $, one can obtain the following expansions~\cite{Cai:2021mrw}:
\beq
\tilde{T}^{(0)}_{\mu \nu} &=& -  \frac{e^{-2A}  }{3} \eta_{\mu \nu} \left( 2 V(\phi_0) + \sum_i \lambda_i (\phi_0) \delta(y - y_i) \right)
\nonumber  \\
\tilde{T}^{(0)}_{55} &=&  \phi'^2_0   +\frac{2}{3} \left( V(\phi_0) + 2  \sum_i \lambda_i (\phi_0)  \delta(y-y_i) \right)  \,\label{T0} 
\nonumber \\
\tilde{T}^{m (0)}_m &=& -\phi'^2_0  -\frac{10}{3} V(\phi_0) -\frac{8}{3} \sum_i \lambda_i  (\phi_0)\delta(y-y_i) \, \label{TT0}
\eeq
\beq
\tilde{T}^{(1)}_{\mu \nu} &=& -\frac{2 e^{-2A}}{3} \left[ \eta_{\mu \nu} \left( \frac{\partial V}{\partial \phi_0} \varphi - 2 V F  \right) +  2 \epsilon \partial_\mu \partial_\nu f(x) V \right] \nonumber \\
&-& \frac{e^{-2A}}{3}  \sum_i \eta_{\mu \nu} \left( \frac{\partial \lambda_i}{ \partial \phi_0} \varphi - \lambda_i (G+ 2F) \right) \delta( y-y_i) \nonumber \\ 
&-&  \frac{2 e^{-2A}}{3}  \sum_i   \lambda_i \, \epsilon \partial_\mu \partial_\nu f(x)  \delta( y-y_i)  \, \label{T1}
\nonumber \\
\tilde{T}^{(1)}_{55} &=& 2 \phi'_0 \varphi'  + \frac{2}{3} \Big[   \sum_i 2 \Big( \lambda_i G  +   \frac{\partial \lambda_i}{ \partial \phi_0} \varphi \Big)\delta(y-y_i) + 2 V G + \frac{\partial V}{\partial \phi_0} \varphi  \Big] 
\eeq
Note that  although the $\epsilon$ is kept in $\tilde{T}_{IJ}^{(1)}$,  one can anticipate the tadpole term does not  depend on the relic gauge.  Substituting $\kappa^2 \times$Eqs.(\ref{T0}-\ref{T1}) into Eq.(\ref{Act}), we will first arrive:
\beq
 \mathcal{L}_{tad} &= & \frac{2}{3} \oint d y  e^{-4A}  \left[ \left(G -4 F\right)+ \epsilon \Yfund f(x) \right]   V(\phi_0)   \nonumber \\ 
& -  &  \frac{1}{3}  \oint d y e^{-4 A} \sum_i  \lambda_i (\phi_0) \left(4 F  -\epsilon \Yfund f(x) \right)  \delta(y-y_i)  \nonumber \\
& + & \frac{1}{3} \oint d y  e^{-4A}   \varphi   \left( 2 \frac{\partial V }{\partial \phi_0} + \sum_i \frac{ \partial \lambda_i }{\partial \phi_0} \delta( y- y_i)  \right) \, \label{tad1}
\eeq
where the terms proportional to  $\phi'^2_0$ and $\phi' \varphi'$ are  cancelled in the linear order expansion. Applying Eqs.(\ref{BG1}-\ref{BG2}) and Eq.(\ref{orth1}),  we can  perform the  transformation:
\beq
&&\frac{2}{3} \oint d y  e^{-4A}  \left[ \left(G -4 F \right)   V  -  \sum_i  2 \lambda_i  F  \delta(y-y_i) \right] \nonumber \\
&=& \frac{1}{3} \oint d y e^{-4A} \left[ G \phi'^2_0 + 4 A' \phi'_0 \varphi \right] \,. \label{tda}
\eeq
Then substituting  Eq.(\ref{tda})   into Eq.(\ref{tad1}),  the radion tadpole becomes:
\beq
 \mathcal{L}_{tad} &= & \frac{1}{3} \oint d y  e^{-4A}  \left[G  \phi'^2_0 +  \left(4 A'  \phi'_0  + \frac{\partial V}{\partial \phi_0}\right) \varphi \right]   \nonumber \\ 
 &+& \frac{1}{3} \oint dy e^{-4A}  \left[ \frac{\partial V}{\partial \phi_0}  + \sum_i  \frac{\partial \lambda_i}{ \partial \phi_0}  \delta(y-y_i) \right] \varphi   \, \label{tad2} \\
 &+& \frac{1}{3} \oint dy e^{-4A} \left[2 V  + \sum_i \lambda_i \delta(y-y_i) \right] \epsilon \Yfund f(x) \nonumber
\eeq
Now we apply the Eqs.(\ref{eom1},\ref{orth1}) and BG equation (\ref{BG0}), the first two lines in Eq.(\ref{tad2}) can be simplified to be:
\beq
&& \frac{1}{\kappa^2} \oint dy \left[ \frac{\kappa^2}{3} \frac{d}{dy} \left( \phi'_0 \varphi e^{-4A}  \right) - e^{-2A} \Yfund \left( F - \frac{A' \epsilon' f(x) }{e^{2A}}  \right) \right] \nonumber \\
& = & \frac{m^2}{\kappa^2} \oint  dy \, e^{-4A}   \left[F   e^{2A} - A' \epsilon' f(x) \right]  \, \label{tdb}
\eeq
where the integration of the total differential term is zero. And again using the BG equations (\ref{BG1}-\ref{BG2}), the third line in Eq.(\ref{tad2}) is rewritten as:
\beq
&&  \frac{1}{\kappa^2} \oint  dy \left[ A' \frac{d}{dy} \left( \epsilon \Yfund f(x)  e^{-4A}  \right) - e^{-4A}  A' \epsilon'  \Yfund f(x) \right]  \nonumber \\
&+& \frac{1}{3} \oint dy \, e^{-4A}  \left[ \phi'^2_0  + \sum_i \lambda_i \delta(y- y_i) \right] \epsilon \Yfund f(x)  \nonumber \\
&=& \frac{m^2}{\kappa^2} \oint dy \, e^{-4A}   A' \epsilon'  f(x)  \, \label{tdc}
\eeq
Combining Eq.(\ref{tdb}) and Eq.(\ref{tdc}), the final expression for the radion tadpole reads:
\beq
 \mathcal{L}_{tad}  &= &  \frac{m^2}{\kappa^2}  \oint dy  \, e^{-4A}  \left[  \left[F   e^{2A} - A' \epsilon' f(x) \right]+ A' \epsilon'  f(x)  \right] \nonumber \\ 
 &= &   \frac{m^2}{\kappa^2}  \oint dy  \,e^{-2A} \, F    \label{tad3}
\eeq
Impressively  the tadpole term  is proportional to  $m^2 e^{-2A} F$ after the  stabilization, with  other spin-0 perturbations eliminated by the EOM and BG equations.  Further investigation requires the knowledge of the  5d profile by solving the radion EOM with proper boundary conditions. In the limit of small back-reaction, the  mass squared of radion is parameterized as $m^2 = \tilde{m}^2  l^2 $  with $l = \frac{\kappa}{\sqrt{2}} \phi_0|_{y=0} $. Therefore to evaluate the tadpole term at the $\mathcal{O}(l^2)$ order,  substituting  the zeroth order profile i.e. $F = c \, e^{2A}$ into Eq.(\ref{tad3}),  we find that the tadpole of the lowest mode  automatically vanishes. While  if the solution of $F$ contains a $\epsilon' $ part as shown  in~\cite{Kogan:2001qx},  Eq.(\ref{tad3}) is  nonzero in general at  the $\mathcal{O}(l^2)$ order.   

\section{Radion Stabilization}
As argued in the previous sections,  only one dynamical  degree of freedom exists for the radion in an $N$-brane  RS model governed by the  action Eq.(\ref{Act}). In fact our analysis is consistent with the naive counting of degree of freedom, since  the 5d  metric  contains $5 = 1_{0} \oplus 2_{\pm 1} \oplus 2_{\pm 2}$  dynamical fields, where the spin-0 scalar  $1_{0}$   plays the role of radion.   A direct consequence of one radion is that just the UV-IR brane distance $L$ is stabilized by the potential.  Expanding the 5d action at the zeroth order,  its derivative with respect to $L$ is exactly zero if Eq.(2.10) holds true and the coordinates  $r_a $ of  intermediate branes  are fixed.  This implies  that the intermediate branes are not dynamical, thus will not generate additional radion-like modes.
In such a way,  the  GW mechanism is generalized into an $N$-brane setup.  For $N= 4$, we will choose the following superpotential for stabilization: 
\beq
W(\phi) = \begin{cases} \frac{6 k_1}{\kappa^2} - u_1 \phi^2 \,, &  0 < y < r  \\[0.3cm]    
\frac{6 k_2 }{\kappa^2} - u_2 \phi^2  \,,  & r < y < L  
\end{cases}    \, \label{pot}
\eeq
where the discontinuity in the first term originates from $\Lambda_1 \neq \Lambda_2$ and the mass parameters $u_{1,2}$ are assumed to be unequal at first.  By matching the singular terms  in Eq.(\ref{eom2}-\ref{eom3}) and Eq.(\ref{BG0}-\ref{BG1}),  the boundary conditions (BC)  with  $\epsilon' =0$ and $G = 2 F$ are derived as:
\beq
&& \left[ F' \right]_i  -  2  \left[A'\right]_i F = \frac{\kappa^2}{3} \left[\phi'_0 \right]_i \varphi  \,, \label{BF}  \\
&& \left[\varphi' \right]_i - 2 \left[\phi'_0\right]_i F = \frac{\partial^2 \lambda_i}{\partial \phi^2}  \varphi \,. \label{Bph}
\eeq
Similar to RS1, we can  impose a stiff potential  at  the UV and  IR branes, leading to $\varphi(y)|_{y =\{ 0, L\}} =0$.  But due to  a single radion field, one can set  $\frac{\partial^2 \lambda_r}{\partial \phi^2} =0$ at $y=  r$ and this will result in a constraint on $u_{1,2}$.  Note that Eq.(\ref{Bph}) corresponds to the BC of  scalar EOM,  hence needs to be consistent with the radion EOM. The left hand of Eq.(\ref{Bph}) can be transformed from Eq.(\ref{eom1a})  to be:
\beq
 \left[\varphi' \right]_r - 2 \left[\phi'_0\right]_r F = \left[ \frac{\phi''_0}{\phi'_0} \right]_r \varphi + \frac{3 e^{2A}}{\kappa^2}  \Box F  \left[\frac{1}{\phi'_0} \right]_r \, \label{Bph1}
\eeq
Using the specific superpotential in Eq.(\ref{pot}),  one can derive the relevant jumps at  $y=r$:
\beq
\left[ \frac{\phi''_0}{\phi'_0} \right]_r = u_2 - u_1 \,, \quad  \, \left[\frac{1}{\phi'_0} \right]_r = \frac{1}{\phi_0(r)} \left(\frac{1}{u_1} -\frac{1}{u_2} \right) \,. \label{ph0}
\eeq
Then substituting Eq.(\ref{ph0}) into Eq.(\ref{Bph1}),  the junction condition becomes:
\beq
 (u_1 - u_2) \left( \phi_0 \varphi - \frac{3}{\kappa^2} \frac{1}{u_1 u_2} e^{2A} \Box F\right)\Big{|}_{y=r}  = 0 \,, \label{Bph2}
\eeq
A trivial solution that satisfies Eq.(\ref{Bph2}) is $u_1= u_2$.  With this choice $\lambda_{\pm r}$ gets no $\phi$ dependence, i.e.$\frac{\partial \lambda_r}{\partial \phi} = \left[ \phi'_0 \right]_r =0$,  and  Eq.(\ref{BF}) determines the BC at $y=r$ to be $\left[ F' \right]_r  =  2  \left[A'\right]_r F $.   Thus by solving  the EOM (\ref{eom})  with the prescribed BC,  one can obtain a  stabilized radion  with its mass below the cutoff scale of IR brane~\cite{Cai:2021mrw}.  

\section{Conclusion}
The main goal of this paper  serves to clarify the  degree of freedom in an $N$-brane  RS model.  First of all,  we provided a complete calculation of the effective Lagrangian till the quadratic order.  Practicing the variation principle in the EFT  delivers  3 EOMs   and  two gauge fixings (\ref{orth1}-\ref{orth2}) originate from the non-mixing condition for the graviton and radion.  Note that Eq.(\ref{orth2}) should be enforced  such that  an exact correspondence between the first two EOMs and the linearized  Einstein equation can be established. Moreover, we illustrated that the scalar EOM removing away brane terms can be derived from the linearized Einstein equation and a single EOM is actually independent. Thus  only one  dynamical degree freedom is allowed  for the radion in an $N$-brane RS model. For consistency,  we  investigate  the linear order expansion in  the effective Lagrangian and  proved that the radion tadpole vanishes at the $\mathcal{O}(l^2)$ order. Hence  the radion EOM is valid  to be derived from the quadratic expansion, without including the tadpole effect.

This paper also clearly explains whether $\epsilon\, \partial_\mu \partial_\nu f(x)$ can be added in $g_{\mu \nu}$ as a radion perturbation.  Without the radion stabilization,  the $\zeta$-symmetry (relic of 5d diffeomorphism) in Eq.(\ref{transform}), keeping the Einstein-Hilbert action  invariant, can remove the $\epsilon$ perturbation in the bulk. By relaxing one BC at an intermediate brane $y= r_a$,  at most a surface kinetic term is induced by $\epsilon'(r_a)$ in the 5d action. However in the scenario that the GW scalar develops a y-dependent VEV, the  $\zeta$-symmetry needs to be  broken in order to preserve the 4d Poincar\'e  symmetry. This symmetry analysis is universal  for an $N$-brane RS model and excludes  the possibility to use $\epsilon'(r_a) \neq 0$ at the intermediate branes  to  gain another radion excitation as proposed by  \cite{Kogan:2001qx}.  In the presence of radion stabilization,  the arbitrariness of $\epsilon(y)$  will cause ambiguity in the radion  kinetic term  \cite{Cai:2021mrw} and  physical observables,  e.g.  the quartic interaction  of  $\epsilon  f(x)$ coupling with one radion  (or one graviton $h_{\mu \nu}$) plus  two SM particles off the branes is not zero in general.  Then $\epsilon'(y)$  has to be  zero  in the bulk and  branes  due to the  5d $\zeta$-symmetry breaking.

\section*{Acknowledgments}
H.C. \ is supported by the National Research Foundation of Korea (NRF) grant funded by the Korea government (MEST) (No. NRF-2021R1A2C1005615).

\newpage
\appendix

\section{Quadratic  expansion of  the 5d  Action}~\label{Appendix1}
In this section, we provide the intermediate steps to derive the  effective Lagrangian  at the quadratic order Eq.(\ref{eff}) except for the $\mathcal{L}_{5m}$ term. 

\subsection{ The  kinetic and $\partial_5^2$ terms of graviton }
First of all we will  calculate  the kinetic  and  $\partial_5^2$ terms of the graviton.  We expand the Ricci tensor till the second order  i.e. $R_{IJ} = R_{IJ}^{(0)}+ R_{IJ}^{(1)} + R_{IJ}^{(2)}$, with the number  inside a pair of parenthesis representing the expansion order. The  zeroth order part is:
\beq
R_{\mu \nu}^{(0)} =   e^{-2A} \, \eta_{\mu \nu}  \left( 4 A'^2 -A'' \right) \,,       \quad  R_{55}^{(0)} = 4 \left( A'' - A'^2  \right) \, \label{R0}
\eeq
And the parts at the higher order including only the $h_{\mu \nu}$ terms read:
 \beq
R_{\mu \nu}^{(1)} &\supset& \frac{1}{2} \left( \partial_\mu \partial_\lambda   
h^{\lambda}_{\nu}  +  \partial_\nu  \partial_\lambda h^{\lambda}_{\mu}    
-  \Yfund h_{\mu \nu} -  \partial_\mu  \partial_\nu h  \right) \nonumber \\
&+& \frac{1}{2} e^{-2 A} \left( \partial_5^2 h _{\mu \nu} - 4 A'  \partial_5 h_{\mu \nu}  \right) \nonumber \\
&+&  e^{-2 A} \left[ 4 A'^2 - A'' \right] h_{\mu \nu} - \frac{1}{2} e^{-2 A} \eta_{\mu \nu}A' \partial_5 h 
\\
R_{55}^{(1)} &\supset&- \frac{1}{2} \left( \partial_5^2 h  - 2 A' \partial_5 h  \right) 
\eeq
\beq
R_{\mu \nu}^{(2)} &\supset & \frac{1}{2} h^{\alpha \beta} \partial_\mu \partial_\nu h_{\alpha \beta} -\frac{1}{2}  \partial_\alpha \left( h^{\alpha \beta}  \left( \partial_\mu h_{\nu \beta} + \partial_\nu h_{\mu \beta} -\partial_\beta h_{\mu \nu} \right) \right) + \frac{1}{4}  \partial_\mu h_{\alpha \beta } \partial_\nu h^{\alpha \beta} \nonumber \\ 
&+&  \frac{1}{2} \partial^\alpha h^\beta_\nu \left( \partial_\alpha h_{\beta \mu} -\partial_{\beta } h_{\alpha \mu }\right)  + \frac{1}{4} \left( \partial^\alpha h\right) \left( \partial_\mu h_{\nu \alpha} + \partial_\nu h_{\mu \alpha} - \partial_\alpha h_{\mu \nu} \right)  \nonumber \\
&+ &\frac{1}{4} e^{-2A} \left[ \eta_{\mu \nu} \left(2 A' h^{ij} \partial_5 h_{ji}\right) + \left( \partial_5 h_{\mu \nu}-2 A' h_{\mu \nu} \right) \partial_5 h - 2 \partial_5 h_{\mu \alpha} \partial_5 h^\alpha_\nu  \right]
\\
R_{55}^{(2)} &\supset& -\frac{1}{4} \partial_5 h^{ij} \partial_5 h_{ij} -A' h^{i \beta} \partial_5 h_{\beta i} +\frac{1}{2} \partial_5 \left( h^{ij} \partial_5 h_{ji} \right) \, \label{R} \label{R2}
\eeq
Note that the linear order expansion was given in \cite{Cai:2021mrw}. With Eq.(\ref{R0}-\ref{R2}), the Ricci scalar  can be calculated at each order, e.g. at the second order:
\beq
R^{(2)} = g^{IJ (0)} R_{IJ}^{(2)} + g^{IJ (1)} R_{IJ}^{(1)} + g^{IJ (2)} R_{IJ}^{(0)} \,. 
\eeq
Some parts of quadratic terms originate from  $ \sqrt{g}^{(0)} R^{(2)} + \sqrt{g}^{(1)} R^{(1)} +  \sqrt{g}^{(2)} \left( R^{(0)} + 2 \kappa^2 V(\phi_0)  \right) $ in  the 5d action Eq.(\ref{Act}):
\beq
\sqrt{g}^{(0)} R^{(2)} + \sqrt{g}^{(1)} R^{(1)} &\supset& e^{-2A} \left[ {\mathcal L}_{FP} + \partial_\alpha \left(  h_{\mu \nu} \partial^\alpha h^{\mu \nu } -   \partial_\mu \left(h^{\mu \nu}  h^\alpha_\nu \right)  - \frac{1}{2} h \partial_\beta h^{\alpha \beta} -\frac{1}{2} h \partial^\alpha h \right)\right]  \nonumber \\
 &+& e^{-4A} \left[ \frac{1}{4} \left[ \partial_5 h_{\mu \nu} \partial_5 h^{\mu \nu} - (\partial_5 h)^2 \right] + A' h^{ij}  \partial_5 h_{ij}    -\frac{1}{2} A' h \partial_5 h \right] \nonumber \\  &- & \partial_5 \left( e^{-4A} h_{\mu \nu} \partial_5 h^{\mu \nu} \right) + \frac{1}{2} \partial_5 \left( e^{-4A} h \partial_5 h \right) ~\label{h2a}
\eeq
with ${\mathcal L}_{FP}$ standing for the Fierz-Pauli Lagrangian:
\beq
{\mathcal L}_{FP}= \frac{1}{2} \partial_\nu h_{\mu \alpha} \, \partial^\alpha    
h^{\mu \nu} -   \frac{1}{4} \partial_\mu h_{\alpha \beta} \, \partial^\mu h^{\alpha \beta}    
- \frac{1}{2} \partial_\alpha h \, \partial_\beta h^{\alpha \beta} +\frac{1}{4}    
\partial_\alpha h \, \partial^\alpha h  \, \label{FP}
\eeq
and
\beq
 \sqrt{g}^{(2)} \left( R^{(0)} + 2 \kappa^2 V(\phi_0)  \right) \supset  -\frac{1}{4} e^{-4A} \left[ h^{\mu \nu}  h_{\mu \nu} -\frac{1}{2} h^2 \right] \left[8 \left( A'^2 - A''  \right) + \kappa^2 \phi'^2_0 \right] \, \label{h2b} 
\eeq
While the scalar kinetic term and the brane potentials in Eq.(\ref{Act}) contribute as well:
\beq
& - & \sqrt{g}^{(2)} \left( \frac{g^{I J (0)}}{2}  \left( \partial_I \phi \partial_J \phi \right)^{(0)} \right) + \sqrt{g_4}^{(2)}\sum_i \lambda_i \delta(y - y_i )\nonumber \\ &\supset& -\frac{1}{4} e^{-4A} \left[ h^{\mu \nu}  h_{\mu \nu} -\frac{1}{2} h^2 \right] \left(  \frac{1}{2} \phi'^2_0 + \sum_i  \lambda_i \delta(y-y_i) \right) \, \label{h2c}
\eeq
Note that  the  total differential terms with $\partial_\alpha$ or $\partial_5$ in Eq.(\ref{h2a})  vanish after the integration.
Thus calculating the quantity of  $-\frac{1}{2\kappa^2} \int dy \left[\mbox{Eq.(\ref{h2a})}+ \mbox{Eq.(\ref{h2b})}+ 2 \kappa^2 \, \mbox{Eqs.(\ref{h2c})} \right]$,  at the second order the effective Lagrangian contains:
\beq
\mathcal{L}^{(2)}  &\supset&  -\frac{1}{2 \kappa^2} \int dy \left( e^{-2A}  \left[ \mathcal{L}_{FP} - \frac{e^{-2A}}{4} \left[  (\partial_5 h)^2  - \partial_5 h_{\mu \nu} \partial_5 h^{\mu \nu}  \right]   \right] + \mathcal{L}_{h^2} \right) \, \label{grav}
\eeq
with
\beq
 \mathcal{L}_{h^2} & = &  \int dy \left[ e^{-4A} \left[ h^{\mu \nu}  h_{\mu \nu} -\frac{1}{2} h^2 \right]    \left( - \frac{\kappa^2}{2}  \left(\phi'^2_0 +  \sum_i \lambda_i \delta(y-y_i) \right) + 2 A'' \right)  \right. \nonumber \\  &+& \left.  \frac{1}{2} A'  \partial_5 \left( e^{-4A} \left[ h^{\mu \nu} h_{\mu \nu}  - \frac{1}{2}   h^2 \right] \right) \right] =0
\eeq
where  the last term can be applied a partial integration and then  using Eq.(\ref{BG1})  sets $ \mathcal{L}_{h^2}$  to equal zero.

\subsection{The radion and graviton mixing terms}
The four radion fields  $F, G, \varphi$ and $\epsilon(y) \partial_\mu \partial_\nu f(x) $ will mix with the graviton field. The mixing via the Ricci tensor  can  only proceed with two $\partial_\mu$   or two $\partial_5$ derivatives. Due to the conformal flatness of  RS metric,  the $\partial_\mu$ operator can not differentiate the radion and graviton perturbations. Therefore  by replacing  only one graviton field in the Fierz-Pauli Lagrangian $\mathcal{L}_{FP}$ to be $h_{\mu \nu} \to - 2 F  \eta_{\mu \nu}$, $h_{55} \to 2 G \eta_{55}$ and $h \to 2 \left(G -4 F\right) $,  we can obtain the  mixing part:
\beq
-\frac{1}{2 \kappa^2}   \int dy e^{-2A} \mathcal{L}_{FP} &\Rightarrow& -\frac{1}{2 \kappa^2}   \int dy e^{-2A} \Big[  2 F  \partial_\mu \partial_\nu h^{\mu \nu} - 2 F  \Yfund \, h \nonumber \\ & + & (G-4F)  \partial_\alpha \partial_\beta h^{\alpha \beta}  -\Yfund \,h (G-4F)  \Big] \nonumber \\
&=& -\frac{1}{2 \kappa^2}  \int dy e^{-2A }  (G- 2F) \left[ \partial_\mu \partial_\nu h^{\mu \nu} -  \Yfund \, h \right] \, \label{mix1}
\eeq
For the  mixing through two $\partial_5$,  the perturbations of $F$ and $G$  should be treated in different approach. Firstly we can pick the term  $- \frac{1}{4} \left[ (\partial_5 h)^2 - \partial_5 h_{\mu \nu} \partial_5 h^{\mu \nu}  \right]$,  and make a single substitution of  $h_{\mu \nu} \to  2 \left[ \epsilon \partial_\mu \partial_\nu f(x) -  F \eta_{\mu \nu} \right]$ and $h \to 2 \left[ \epsilon \Yfund f(x) - 4 F\right] $ to obtain:
\beq
&&\frac{1}{2 \kappa^2}  \int dy \frac{e^{-4A}}{4} \left[  (\partial_5 h)^2  - \partial_5 h_{\mu \nu} \partial_5 h^{\mu \nu}  \right] \nonumber \\ &\Rightarrow&  \frac{1}{2 \kappa^2}   \int dy e^{-4A}  \left[  \partial_5 ( \epsilon \Yfund f(x) ) \partial_5 h -  \partial_5 ( \epsilon  \partial_\mu \partial_\nu f(x) ) \partial_5 h^{\mu \nu}) \right] \nonumber \\ & - &   \frac{1}{2 \kappa^2}  \int dy e^{-4A}  \left[  \partial_5 (4 F ) \partial_5 h -  (F \eta_{\mu \nu}) \partial_5 h^{\mu \nu}) \right]  \nonumber \\
& = &   \frac{1}{2 \kappa^2} \int dy e^{-4A} \left[ \left(\epsilon' \Yfund f(x) \partial_5 h - \epsilon' \partial_\mu \partial_\nu  f(x)  \partial_5 h^{\mu \nu}\right) - 3  F' \partial_5 h \right] \, \label{mix2}
\eeq
Next the  mixings between $G$, $\varphi$ and $\partial_5 h$ have to be directly calculated.  For clarity, we will list the Ricci tensor that contributes to the  mixing  of $G$ and $ \partial_5 h$ as: 
\beq
R_{\mu \nu}^{(1)} &\supset& e^{-2A} \eta_{\mu \nu} \left[ A' G' - 2G \left(4 A'^2 - A'' \right) \right] \, \label{RG1}
\\
R_{55}^{(1)} &\supset& -\frac{1}{2} \left( \partial_5^2  h - 2 A' \partial_5 h \right) \, \label{Rh1}
\eeq
\beq
R_{\mu \nu}^{(2)} &\supset& e^{-2A} \big[ 3 A' G \left( \partial_5 h_{\mu \nu}-2 A' h_{\mu \nu} \right) + G A'  \left( \partial_5 h \eta_{\mu \nu} - \partial_5 h_{\mu \nu}\right)  \nonumber \\ &-&\frac{1}{2} G' \left( \partial_5 h_{\mu \nu} - 2A' h_{\mu \nu}\right)  \nonumber \\  & - & G\partial_5 \left( \partial_5 h_{\mu \nu} - 2 A' h_{\mu \nu}\right)  -2 A'^2 G h_{\mu \nu} \big] \, \label{RGh1}
\\
R_{55}^{(2)} &\supset& \frac{1}{2} G' \partial_5 h \label{RGh2}
\eeq
Combining Eqs.(\ref{RG1}-\ref{RGh2}), we can extract the relevant terms in  $\sqrt{g}^{(0)} R^{(2)} + \sqrt{g}^{(1)} R^{(1)}$:
\beq
&& \sqrt{g}^{(0)} R^{(2)} + \sqrt{g}^{(1)} R^{(1)} \nonumber \\ &&\supset e^{-4A} \left[ -G \partial_5^2 h + 5 G A' \partial_5 h - G' \partial_5 h + 4 A' G'  h - 4 Gh \left( 5 A'^2 - 2 A''\right) \right] \, \label{RGh3}
 \eeq
 There are  similar terms  originating from  the expansion of $ \sqrt{g}^{(2)} \left( R^{(0)} + 2 \kappa^2 V(\phi_0)  \right)$:
 \beq
 \sqrt{g}^{(2)} \left( R^{(0)} + 2 \kappa^2 V(\phi_0)  \right) &\supset&  \frac{1}{2} e^{-4A}G h  \left[8 \left( A'^2 - A''  \right) + \kappa^2 \phi'^2_0 \right] \, \label{RGh4}
\eeq
Then we  add the contribution from the scalar kinetic term: 
\beq
- \frac{1}{2} \left( \sqrt{g}^{(2)}  g^{I J (0)} + \sqrt{g}^{(1)}  g^{I J (1)}  \right)  \left( \partial_I \phi \partial_J \phi  \right)^{(0)} \supset - \frac{1}{4}  e^{-4A} G h   \phi'^2_0 \label{RGh5}
\eeq
Finally  substituting Eqs.(\ref{RGh3}-\ref{RGh5}) into Eq.(\ref{Act}),  we find that the terms proportional to $\phi'^2_0$  are all cancelled,  and the mixing term of   $G \partial_5 h$  is derived to be:
\beq
\mathcal{L}_{Gh} &=& -\frac{1}{2 \kappa^2 } \int dy \left[ -3  e^{-4A}  G A' \partial_5 h  -\partial_5 \left( e^{-4A} G \partial_5 h \right) \right] \nonumber \\
&-& \frac{2}{\kappa^2 } \int dy \left[  e^{-4A} G h A''   +   A' \partial_5 \left( e^{-4A} G h \right) \right] \nonumber \\
&=&  \frac{3}{2 \kappa^2 } \int dy   e^{-4A}  G A' \partial_5 h \, \label{mix3}
\eeq
Now we will show how to calculate the  mixing term involving  $\varphi \partial_5 h$. Such type of  term  comes from the following combination:
\beq
&- & \int dy \left[- \sqrt{g}^{(1)} \left(\frac{g^{IJ (0)}}{2}  \left( \partial_I \phi \partial_J \phi \right)^{(1)} \right)+ \sqrt{g}^{(1)}   V^{(1)} + \sum_i \sqrt{g_{4}}^{(1)} \lambda_i^{(1)} \delta(y-y_i) \right] \nonumber \\
 &\supset&  -  \int dy e^{-4A}  \frac{1}{2}  h \left[ \phi'_0 \varphi' + \frac{\partial V}{\partial \phi_0} \varphi +  \sum_i \frac{\partial \lambda_i}{\partial \phi_0} \varphi  \delta(y-y_i) \right] \nonumber  \\
 &=& -\int dy e^{-4A} \frac{1}{2}  h \left[ \partial_5 \left( \phi'_0 \varphi \right) - \left( \phi''_0   - \frac{\partial V}{\partial \phi_0}  -  \sum_i \frac{\partial \lambda_i}{\partial \phi_0}\delta(y-y_i) \right) \varphi    \right] \nonumber \\
 &=& -\int dy e^{-4A} \frac{1}{2}  h \left[ \partial_5 \left( \phi'_0 \varphi \right) - 4 A' \phi'_0 \varphi \right]  =   \frac{1}{2}   \int dy   e^{-4A}\phi'_0 \varphi  \partial_5 h \, \label{mix4}
\eeq
Adding up Eqs(\ref{mix1}-\ref{mix2}) and Eqs.(\ref{mix3}-\ref{mix4}), we obtain  the final forms of the mixing between the radion and graviton:
\beq
\mathcal{L}_{mix}  &=& -\frac{1}{2 \kappa^2} \int dy e^{-2A}\left[ \left[ G- 2 F - e^{2A}  \partial_5 \left( \epsilon'  f(x) e^{-4A}\right) \right] \left( \partial_{\mu} \partial_\nu h^{\mu \nu} - \Yfund \, h\right) \right. \nonumber \\ &+&  \left. 3 e^{- 2A} \left[ F'  -A' G  - \frac{\kappa^2}{3} \phi'_0 \varphi \right] \partial_5  h  \right] \, \label{mix} 
\eeq

\subsection{The kinetic term of radion}
The kinetic term of radion contains three parts: 1) involving only $F$ and $G$ perturbations,  2) with one $\epsilon' \Yfund f(x)$, and 3) involving only the scalar  perturbation $\varphi$. For the first part, we can derive it from the kinetic term of graviton $-\frac{1}{2 \kappa^2}   \int dy e^{-2A} \mathcal{L}_{FP}$ by replacing  $h_{\mu \nu} \to - 2 F  \eta_{\mu \nu}$, $h_{55} \to 2 G \eta_{55}$ and $h \to 2 \left(G -4 F\right) $:
\beq
 \mathcal{L}_{rad} &\supset& - \frac{1}{2 \kappa^2}   \int dy e^{-2A} \left[ 2 \partial_\mu  F \partial^\mu F - \left( 4 \partial_\mu F \partial^\mu F + \partial_\mu G \partial^\mu G \right) \right. \nonumber \\ &&   +   \left.  2 \partial_\mu \left( G - 4F\right) \partial^\mu  F + \partial_5 (G- 4 F) \partial^\mu \left(G- 4F\right)\right] \nonumber \\
 &=& -\frac{3}{\kappa^2} \int dy e^{-2A} \partial_\mu F \partial^\mu \left( F- G \right) \, \label{kin1}
\eeq
And the second part  can be calculated  from  Eq.(\ref{mix}) by substituting $h_{\mu \nu} \to 2 \epsilon \partial_\mu \partial_\nu f(x) $ and   $h \to 2 \epsilon  \Yfund f(x)$:
\beq
\mathcal{L}_{rad}  &\supset& -\frac{3}{\kappa^2} \int dy e^{-4A}  \left[ F'  -A' G  - \frac{\kappa^2}{3} \phi'_0 \varphi \right] \epsilon' \partial_\mu \partial^\mu f(x)  \nonumber \\
&=&  \frac{3}{\kappa^2} \int dy e^{-4A} \epsilon'  \partial_\mu \left[ F'  -A' G  - \frac{\kappa^2}{3} \phi'_0 \varphi \right]  \partial^\mu f(x) + \mbox{the surface term}\, \label{kin2}
\eeq
Combining Eqs. (\ref{kin1}-\ref{kin2}) and the  part 3), the total expression for the kinetic term of radion field is:
\beq
 \mathcal{L}_{rad} &=& \frac{1}{2} \int dy e^{-2A} \left[ \partial_\mu \varphi \partial^\mu \varphi -\frac{6}{\kappa^2}  \Big[\partial_\mu F \partial^\mu \left( F- G \right)  \right. \nonumber \\ &-& \left. e^{-2A} \epsilon'  \partial_\mu \left[ F'  -A' G  - \frac{\kappa^2}{3} \phi'_0 \varphi \right]  \partial^\mu f(x) \Big] \right] \, \label{rad}
\eeq

\section{Correlation of  the scalar  EOM and  Einstein Equation}~\label{Appendix3}
The  scalar EOM (\ref{eom3})  can be derived  from  the linearized Einstein equations.  Analogously  the  $(\mu \nu)$ and $(55)$ components of Eq.(\ref{EE}) are not independent.  We  start with  providing the detail to show the correlation between the scalar EOM and Einstein equation.
Firstly we need to evaluate  $ \partial_5 \left[e^{-2A}  \mbox{Eq.(\ref{eom})}\right]$. By transforming Eq.(\ref{eom}) back into Eq.(\ref{eom1a}), this gives:
\beq
&&\Box F'  - \partial_5 \left(e^{-2A} \left(A' \epsilon' \Box f(x) -  \frac{\kappa^2}{3} \phi'^2_0 G \right)\right)  \nonumber \\ &-&\frac{\kappa^2}{3} e^{-2A} \left( \phi'_0 \varphi'' - \phi'''_0 \varphi -2 A' \left(\phi'_0 \varphi' - \phi''_0 \varphi \right) \right) = 0 \, \label{dEOM} 
\eeq
Dropping the boundary terms,   $A' e^{-2A} \left(R_{55} - \kappa^2 \tilde{T}_{55}\right) =0$ gives~\cite{Cai:2021mrw}:
\beq
&&\left[ - (\epsilon'' - 2 A' \epsilon') \Box f(x)  + e^{2A}  \Box G  + 4 F''  - 4 A' (G' + 2 F') \right] A' e^{-2A} \nonumber \\
&=& \kappa^2 \left[ \frac{4}{3} G V + 2 \phi'_0 \varphi' + \frac{2}{3} \frac{\partial V}{\partial \phi} \varphi  \right] A' e^{-2 A}\, \label{R55}
\eeq
Now one can immediately calculate  the quantity of (Eq.(\ref{dEOM}) $-$ Eq.(\ref{R55})) to be:
\beq
&& \Box \left( F'- A' G\right) - e^{-2A} A'' \epsilon' \Box f(x)  + \frac{ \kappa^2}{3} \phi'_0 \left( 4  A'  \varphi' -  \varphi'' e^{-2A} + \frac{\partial^2 V}{\partial \phi^2_0} \varphi   \right) e^{-2A}  \nonumber \\ &+&  \left[\frac{\kappa^2}{3} \left(2 \phi'_0 \phi''_0 G + \phi'^2_0 G'  + 4 A'' \phi'_0 \varphi  \right)  - 4 A' A'' G \right] e^{-2A} \nonumber \\
&+ & 8  \left[ \frac{ \kappa^2}{6} G A'  \left( V - \frac{1}{2} \phi'^2_0  \right)  +  A'^2 F'  \right] e^{-2A}+ \frac{2 \kappa^2}{3} \left( \frac{\partial V}{\partial \phi} - \phi''_0 \right) \varphi A' e^{-2A}  =0 \label{scalar}
\eeq
where the first two term can be combined as:  
\beq
 \Box \left( F'- A' G\right) - e^{-2A} A'' \epsilon' \Box f(x) =  \frac{\kappa^2}{3} \phi'_0 e^{-2A} \Box \left( \varphi  e^{2A}-  \phi'_0 \epsilon' f(x) \right) \, \label{dE1}
\eeq
and  applying Eqs.(\ref{BG0}-\ref{BG1}) for the terms in the second line,  we can rewrite:
\beq
&& \left[\frac{\kappa^2}{3} \left( 2 \phi'_0 \phi''_0 G + \phi'^2_0 G'  + 4 A'' \phi'_0 \varphi  \right) - 4 A' A'' G \right] e^{-2A}  \nonumber \\
&=&  \frac{\kappa^2}{3}\phi'_0 \left[ 2 \frac{\partial V}{\partial \phi}  G  + \phi'_0 ( G' + 4F') \right] e^{-2A} \, \label{dE2}
\eeq
Then using Eq.(\ref{BG2}) and Eq.(\ref{orth1}), one can obtain: 
\beq
\left[ \frac{ \kappa^2}{6} G A'  \left( V - \frac{1}{2} \phi'^2_0  \right)  +  A'^2 F'  \right] e^{-2A} =  \frac{ \kappa^2}{3} A'^2 \phi'_0 \varphi e^{-2A} \, \label{dE3}
\eeq
Combining  Eq.(\ref{dE3}) with the last term in Eq.(\ref{scalar}) and applying Eq.(\ref{BG0}),  we can find that all the terms in the third line of Eq.(\ref{scalar}) are exactly cancelled. Finally substituting Eqs.(\ref{dE1}-\ref{dE2}) into Eq.(\ref{scalar}),  that equation reproduces the scalar EOM (\ref{eom3}) times $\kappa^2\phi'_0 e^{-2A}/3 $.

Next we will prove that the $(\mu \nu)$ part of Eq.(\ref{EE}) can be derived from its  $(55)$ part.  Expanding to the linear  order,  one can extract out the $\eta_{\mu \nu}$ part in $R_{\mu \nu} = \kappa^2 \tilde{T}_{\mu \nu}$~\cite{Cai:2021mrw}:
\beq
&& e^{2A} \Box \left[ F - A' \epsilon' e^{-2A} f(x) \right] - F'' + A' (8 F' +G')  \nonumber \\
&=& -\frac{\kappa^2}{3} \left[ 4 G V + 2 \frac{\partial V}{\partial \phi_0} \varphi + \sum_i \left( \frac{\partial \lambda_i}{\partial \phi_0} \varphi+ \lambda_i G \right) \delta(y-y_i) \right] \label{Euv}
\eeq
as the $\partial_\mu \partial_\nu$ part is a gauge fixing. On the other hand,  the linearized $R_{55} = \kappa^2 \tilde{T}_{55}$ is~\cite{Cai:2021mrw}:
\beq
&& e^{2A} \Box \left[  G - \left( \epsilon'' - 2A' \epsilon' \right) e^{-2A} f(x) \right]    + 4 F'' - 4 A' (G'+ 2F') \nonumber \\
&=& \frac{2 \kappa^2}{3} \left[ 2 GV  +  \frac{\partial V}{\partial \phi_0} \varphi  +  3 \phi'_0 \varphi'  + 2 \sum_i \left( \frac{\partial \lambda_i}{\partial \phi_0} \varphi+  \lambda_i G  \right)\right] \,, \label{E55}
\eeq
The derivative of  the first orthogonal condition Eq.(\ref{orth1}) gives:
\beq
\phi'_0 \phi' = - \phi''_0 \phi +  \frac{3}{\kappa^2} \left( F'' - A' G' - A'' G \right) \label{dorth1}
\eeq
By substituting Eq.(\ref{dorth1}) and the second orthogonal condition Eq.(\ref{orth2}) into Eq.(\ref{E55}), one can arrive that:
\beq
&&  e^{2A} \Box \left( F  - A' \epsilon' e^{-2A} f(x) \right)  -   F''  + A' (8 F' +G')+ 3 (A'' - 4 A'^2 ) G \nonumber \\
&=& \frac{\kappa^2}{3} \left[ 2 G V -  2 \frac{\partial V}{\partial \phi_0} \varphi  - \sum_i \left(  \frac{\partial \lambda_i}{\partial \phi_0} \varphi - 2 \lambda_i G  \right) \delta(y - y_i )\right]  \, \label{Euv0}
\eeq
Then applying Eq.(\ref{BG1}-\ref{BG2}) to  Eq.(\ref{Euv0}),  the $(55)$ part of Einstein equation  is explicitly transformed into the $(\mu\nu)$ one   in  Eq.(\ref{Euv}).

\section{Proof of sum rule}~\label{Appendix2}  
The  absence of tadpole was claimed in the literature as the consequence of the following ansatz:
\beq
\oint dy e^{-2 A} T^{\mu}_\mu =  \oint dy e^{-2A} \left[  -2 \phi'^2_0 -4 V(\phi_0) -4 \sum_i \lambda_i \delta(y- y_i) \right] =  0 \,, \label{sumrule0}
\eeq
and we are going to provide an alternative proof for this sum rule. Due to $\phi'_0 =  \frac{1}{2}   \frac{\partial W(\phi_0)}{\partial \phi_0}$ in Eq.(\ref{sup}) and counting  the discontinuity of the superpotential $W(\phi)$ at the junctions, one gets:
\beq
\phi'^2_0 = \frac{1}{2}   \frac{\partial W}{\partial \phi_0}  \frac{\partial \phi_0}{\partial y} =  \frac{1}{2} \left[  \frac{\partial W}{\partial y}  - 2 \sum_i \lambda_i \delta(y-y_i)  \right] \,\label{phi0}
\eeq
Applying $V(\phi) =  \frac{1}{8} \left(\frac{ \partial W(\phi)}{\partial \phi}\right)^2 -\frac{\kappa^2}{6} W(\phi)^2$ in Eq.(\ref{sup}), we can rewrite Eq.(\ref{sumrule0}) in the following form:
\beq
&& \oint dy e^{-2A} \left[  -2 \phi'^2_0 -4 V(\phi_0) -4 \sum_i \lambda_i \delta(y- y_i) \right] \nonumber \\
 &=&  \oint dy e^{-2A} \left[  -2 \phi'^2_0 - \frac{1}{2}  \left( \frac{\partial W (\phi_0)}{\partial \phi_0} \right)^2 + \frac{2 \kappa^2}{3} W(\phi_0)^2 -4 \sum_i \lambda_i \delta(y- y_i) \right] \nonumber \\
 &=&   \oint dy e^{-2A} \left[  -4 \phi'^2_0  + \frac{2 \kappa^2}{3}  W(\phi_0)^2 -4 \sum_i \lambda_i \delta(y- y_i) \right] 
\eeq
Now we can insert Eq.(\ref{phi0}) to remove the singular term and after partial integration this gives:
\beq
\oint dy e^{-2 A} T^{\mu}_\mu  &= & \oint dy  \left[ e^{-2A} \left( -4 A' W + \frac{2\kappa^2}{3} W^2\right)  - 2 \frac{d}{dy} \left( e^{-2A} W\right) \right] \nonumber \\
&=& \oint dy e^{-2A} \left[ -4 A' W + \frac{2\kappa^2}{3} W^2  \right] =0
\eeq
where the total differential term vanishes and in the last line we used $A'= \frac{\kappa^2}{6} W$ in Eq.(\ref{sup}).

\bibliographystyle{JHEP}
\bibliography{Nbrane}

\end{document}